\documentstyle[epsfig,preprint,aps]{revtex}
\begin{document}
%
\draft
\title{Quarks in nuclear medium}
%
\author{K. Saito}
\address{Tohoku College of Pharmacy, Sendai 981-8558, Japan}
\maketitle
\begin{abstract}
Using the quark-meson coupling (QMC) model we study nuclear matter from
the point of view of quark degrees of freedom.  As the
nucleon model we adopt the MIT bag model and the relativistic constituent
quark model, where a square well and harmonic oscillator
potentials are used to confine the quarks.  We introduce
the Lorentz-vector type confining potential as well as the Lorentz-scalar
type one in order to examine how the vector confining potential contributes 
to the properties of the nucleon and nuclear matter. 
Next, we perform a re-definition of the scalar field in matter and
transform the QMC model to a QHD-type model with a non-linear scalar
potential.  The result obtained from QMC is then compared with
the potentials 
which are determined so as to fit various properties of finite nuclei and
nuclear matter in relativistic mean-field models.  The QMC model
provides the parameters $\kappa \sim 20 - 40$ (fm$^{-1}$) and
$\lambda \sim 80 - 400$ for the standard, non-linear scalar potential.
We discuss a relationship between the QMC and QHD-type models in detail.
\end{abstract}
%

\newpage
\section{Introduction}
\label{sec:intro}

In the conventional nuclear physics we have assumed that the nucleon
properties in the nuclear medium are not changed from those of the free
nucleon.  Thus, low- and medium-energy nuclear physics phenomenology
has been successfully described in terms of (point-like) nucleons.
However, after the EMC effect in the nuclear structure functions was
observed~\cite{emc} there has been considerable interest in the modification
of hadron properties in the nuclear medium.  
It is one of the central questions 
in nuclear physics how the effect of the quark substructure of hadrons
emerges in nuclear phenomena.

At present rigorous studies of quantum chromodynamics (QCD) are limited to
matter system with high
temperature and zero baryon density.  Because of the complex
non-perturbative features of QCD it is very difficult to calculate 
properties of nuclei at low- and medium-energy. 
Thus, we need to build models which help to bridge the discrepancy between
nuclear phenomenology and the underlying theory.  Such models
may be necessarily crude because the nuclear many-body problem at the
fundamental QCD level is intractable.  However, it is very important and
intersting to challenge this problem.  Then, much effort has been devoted
to the study of effective models for nuclear matter and finite nuclei based
on various quark models~\cite{various}.

About a decade ago, Guichon~\cite{guichon} proposed a relativistic
quark model for nuclear matter, where
it consists of non-overlapping nucleon bags bound by the self-consistent
exchange of isoscalar, scalar ($\sigma$) and vector ($\omega$) mesons
in mean-field approximation (MFA).
This model has been further developed as the quark-meson coupling (QMC)
model~\cite{guichon,qmc,qmc2}.  It was shown that the model can reproduce the
saturation properties of nuclear matter and that it provides a fair
description of finite nuclei (for recent reviews, see
Ref.\cite{review1}). 
Since in QMC the mesons couple not to the nucleons but to the 
quarks directly 
the nucleon properties are modified self-consistently, depending
on the scalar field in a medium. 

On the other hand, recent theoretical studies show that various properties
of finite nuclei can be described by Quantum Hadrodynamics (QHD)~\cite{qhd}
very well.  Although the original version of QHD~\cite{qhd} provides
the large reduction of the nucleon mass in nuclear matter, which is the
source of the large spin-orbit splitting in finite nuclei, the model also
gives the large nuclear compressibility around 550 MeV.  In order to
improve this Boguta and Bodmer~\cite{boguta} first introduced 
self-interaction terms for the $\sigma$ meson.  It is now well established
that such non-linear interactions are necessary to reproduce the bulk
properties of finite nulcei and nuclear matter~\cite{review2}.

Our purpose here is to study a relation between QHD-type mean-field models
and the QMC model and to examine how the internal 
structure of the nucleon sheds its effect on effective nuclear models. 
In QMC the nucleon mass in matter is given by a decreasing function of the
$\sigma$ field.  By applying a field re-definition to the scalar 
field in matter, 
the QMC model can be cast in a form equivalent to a QHD-type
model with a non-linear scalar potential and a non-linear coupling to
the gradient of the scalar field~\cite{muller}.  Therefore, 
in principle, if the effective nucleon mass is known as a function of the
scalar field, it could provide a prediction of non-linear interactions of 
the scalar field in nuclear matter through the procedure of the 
field re-definition.  Inversely, it may imply that if
a form of the scalar self-interactions in matter could be determined by
analysing various experimental data we could get significant information
on the quark substructure of the in-medium nucleon. 

The QMC model relies on the choice of a quark model of the nucleon.
The MIT bag model has been used as the nucleon model in our previous
works~\cite{guichon,qmc,qmc2,review1}.  In the bag model
the inside of the bag is assumed to be the perturbative vacuum and the
quark moves with a current quark mass, which is nearly zero.
There is an alternative idea for the nucleon model, namely
the constituent quark model~\cite{const}.  In this model, the quark takes 
a constituent quark mass of about 300 MeV in the nucleon, which is offered
by spontaneous chiral symmetry breaking. 
Recently, Shen and Toki~\cite{toki} have proposed a new version of 
the QMC model, where 
the constituent quark model with a harmonic oscillator potential 
was used, and studied nuclear matter and finite  
nuclei --- they refer the model as the quark mean-field (QMF) 
model.  Although the quark acquires a heavy mass 
in this picture it is still important to treat it relativistically
because the quark kinetic energy is not small compared with the
constituent quark mass.  Hence, in this paper we will use both the bag model 
and the relativistic constituent quark model in QMC and  
study the properties of nuclear matter and self-interactions of the
scalar field. 

The outline of the present paper is as follows: In Sec.~\ref{sec:quark},
we present relativistic quark models for the nucleon, which can be solved
analytically.  Using those quark models 
we calculate the nucleon properties in
Sec.~\ref{sec:qmc}. We also present our numerical results for nuclear matter.
In Sec.~\ref{sec:qhdtype}, by performing a re-definition of the scalar
field we transform the QMC model to a QHD-type model.
Then, we discuss non-linear scalar potentials generated from the QMC model.
In Sec.~\ref{sec:concl} we give some discussions and conclusion.

\section{Quark models for the nucleon}
\label{sec:quark}

In this section we summarize relativistic quark models to describe
the nucleon.  First we discuss the constituent quark model with a square
well (SW) potential or a harmonic oscillator (HO) potential to confine
the quarks.   We introduce a Lorentz-vector type potential as well
as a scalar type one to examine the role of the vector type 
confining potential in nuclear matter.  In the constituent quark model 
it is assumed that the quark moves 
in the Nambu-Goldstone vacuum and hence the quark takes a constituent quark
mass (around 300 MeV).  Next, we review the MIT bag model shortly.
In the bag model, the quark moves in the perturbative vacuum (or the Wigner
phase) and it takes a small, current quark mass.

\subsection{Constituent quark model}
\label{subsec:constituent}

We consider a light (u or d) quark (of mass $m_q$) moving under a potential
$V(r)$.
The Dirac equation for the quark field $\psi_q(r)$ is then given by
\begin{equation}
[i\gamma\cdot\partial - m_q - V(r)] \psi_q(r) = 0,
\label{dirac}
\end{equation}
where we take
\begin{equation}
V(r) = (1 + \beta \gamma_0) U(r),
\label{potential}
\end{equation}
with a confining potential $U(r)$ and a parameter $\beta (\geq 0)$ to
control the strength of the Lorentz-vector type potential.  We assume
that the shape of the Lorentz-scalar type confining potential is the same
as that of the vector type one. 

We first consider a square well potential and find a solution for the
quark field {\it {\`a} la Bogolioubov}~\cite{bogo}.  The potential is then 
given by
\begin{equation}
U(r) = \left\{ {0, \ \ \ \mbox{for} \ r \leq R, \atop
M, \ \ \ \mbox{for} \ r > R,} \right.
\label{swpot}
\end{equation}
where $R$ is the radius of the spherical well and $M$ is
the height of the potential outside the nucleon.
After finishing all calculations, we take
the limit $M \to \infty$.

The solution for the quark field can be found easily~\cite{bogo}.
Inside the well the lowest energy state is
\begin{equation}
\psi_q(r) = \frac{N}{\sqrt{4\pi}} e^{-i Et}
{f(r)\choose i{\vec \sigma}\cdot{\hat r}g(r)} \chi_s,
\label{qsol}
\end{equation}
with the normalization constant $N$, spin state $\chi_s$ and
the quark energy $E$.  Here the upper $f(r)$ and lower $g(r)$ components 
of the quark wave function 
are respectively given by the spherical Bessel functions, $j_0(r)$ and
$j_1(r)$.  (For details, see Appendix.)
Outside the nucleon volume we find a similar form of the solution with
modified spherical Bessel functions.  Note that since the quark wave
function is proportional to $e^{-M\sqrt{1-\beta^2}r}$ outside the well 
the condition $\beta < 1$ is necessary to confine the quark.

We then demand that $f$ and $g$ must be continuous at $r = R$.  This
gives an eigenvalue problem, and we obtain a matching condition in
the limit $M \to \infty$:
\begin{equation}
j_0(x) = \sqrt{\frac{(1-\beta)(E-m_q)}{(1+\beta)(E+m_q)}} j_1(x),
\label{bound}
\end{equation}
with the eigenvalue of the confined quark, $x$.  The total energy
is given by $3 \alpha /R$, where 
$\alpha^2 = x^2+\lambda^2$ and $\lambda = Rm_q$.  As in the MIT bag model
there exist the spurious center of mass (c.m.) motion and gluon fluctuation
corrections etc.  Here we add the familiar form, $-z/R$, 
with a parameter $z$ for those corrections to 
the energy~\cite{bogo}.  Thus, the nucleon mass $M_N$ (at rest) is given by
\begin{equation}
M_N = \frac{3 \alpha - z}{R}.
\label{nmass1}
\end{equation}
Using the solution Eq.(\ref{qsol}) we can calculate
various nucleon properties analytically (see Appendix).

This system may be described by a Lagrangian density
\begin{equation}
{\cal L}_{SW} = {\bar \psi}_q (i\gamma\cdot\partial - m_q) \psi_q
\theta(R-r) - \frac{1}{2}{\bar \psi}_q (1+\beta \gamma\cdot a)
\psi_q \delta(r-R), 
\label{swlag}
\end{equation}
where $a^\mu$ is the unit vector in time direction: $a^\mu =
(1, {\vec 0})$.  This Lagrangian provides a boundary condition at $r=R$
\begin{equation}
i \gamma\cdot n \psi_q = (1+\beta \gamma\cdot a) \psi_q,
\label{bound2}
\end{equation}
where $n^\mu$ is the unit normal outward from the potential surface.
This is equivalent to the condition Eq.(\ref{bound}) and ensures that
the quark is confined permanently.  (Note that actually the quark current
flows in the azimuthal direction~\cite{saito}.)

In the case where the confining potential is a harmonic
oscillator potential with $\beta = 1$ and $c$ the oscillator strength 
\begin{equation}
V(r) = \frac{1}{2}(1 + \gamma_0) c r^2,
\label{hopot}
\end{equation}
one can again find the quark 
wave function analytically.  The solution to the Dirac equation is
expressed by Eq.(\ref{qsol}), where the upper and lower components
are given in terms of gaussian functions~\cite{ho1}.  (For details,
see Appendix.)
The condition to determine the quark energy $E$ is then obtained
\begin{equation}
\sqrt{E+m_q} (E-m_q) = 3 \sqrt{c}.
\label{hocond}
\end{equation}

In the HO model the c.m. energy can be estimated exactly as in the
non-relativistic harmonic oscillator. It is just one third of
the total energy~\cite{ho2}.  Thus, the nucleon mass is
\begin{equation}
M_N = 2E - E_s,
\label{nmass2}
\end{equation}
where $E_s$ describes gluon fluctuation corrections etc.
Note that an equally
weighted (Lorentz) scalar-vector potential does not produce a spin-orbit
splitting in the baryon spectrum, which may be consistent with the fact that 
the spin-orbit splitting in the observed baryon spectra is very small.  

\subsection{MIT bag model}
\label{subsec:MITbag}

In the MIT bag model the inside of the nucleon bag is considered to be
the perturbative vacuum, and the quark takes a current quark mass.
It requires extra energy to make a nucleon bag in which the quark moves
freely, so we have to add a latent heat term to the Lagrangian density
Eq.(\ref{swlag})
\begin{equation}
{\cal L}_{B} = {\cal L}_{SW} - B \theta(R-r),
\label{blag}
\end{equation}
where the bag constant $B$ describes the energy gap between the inside
and outside of the nucleon.

The quark wave function and the eigenvalue condition are given by the same
forms as in the SW model.  Thus, the nucleon mass is given as
\begin{equation}
M_N = \frac{3 \alpha - z}{R} + \frac{4\pi}{3} B R^3,
\label{nmass3}
\end{equation}
where the $z$ parameter is again introduced to take into account the sum of
the c.m. and gluon fluctuation corrections etc.

\section{QMC model for nuclear matter}
\label{sec:qmc}

\subsection{Effect of nucleon structure in a nuclear medium}
\label{subsec:nucleonstr}

Here we consider how Eq.(\ref{dirac}) is modified when the nucleon is bound
in static, uniformly distributed nuclear matter.
In the QMC model~\cite{guichon,qmc,qmc2} it is assumed that
each quark feels scalar $V_s^q$ and vector $V_v^q$ potentials, which are
generated by the
surrounding nucleons, as well as the confinement potential $V(r)$.
Since the typical distance between two nucleons around normal nuclear
matter density 
($\rho_0 = 0.15$ fm$^{-3}$) is surely larger than the typical size of the
nucleon (the radius is about 0.8 fm), the interaction (except for
the short-range part) between the nucleons
should be colour singlet, namely a meson-exchange potential.  Therefore,
this
assumption seems appropriate when the baryon density $\rho_B$ is not high.
(For high density, see Ref.~\cite{high}.)
If we use the mean-field approximation for the meson fields,
Eq.(\ref{dirac}) may be rewritten as
\begin{equation}
[ i\gamma\cdot\partial - (m_q - V_s^q) - V(r)
 - \gamma_0 V_v^q ] \psi_q(r) = 0 .
\label{dirac2}
\end{equation}
The potentials generated by the medium are constants because the matter
distributes uniformly. As the nucleon is static, the time-derivative
operator in the Dirac equation can be
replaced by the quark energy, $-i E$.

By analogy with the procedure applied to the nucleon
in QHD~\cite{qhd}, if we introduce the
effective quark mass by $m_q^{\star} = m_q - V_s^q$, the Dirac equation
Eq.(\ref{dirac2}) can be rewritten in the same form as that in free space,
with the mass $m_q^{\star}$ and the energy $E - V_v^q$, instead of
$m_q$ and $E$.  In other words, the vector interaction has {\em no effect
on the nucleon structure} except for an overall phase in the quark wave
function Eq.(\ref{qsol}), which gives a shift in the nucleon energy.  
This fact 
{\em does not\/} depend on how to choose the confinement potential $V(r)$.
Then, the nucleon energy (at rest) $E_N$ in the medium 
is~\cite{guichon,qmc,qmc2}
\begin{equation}
E_N = M_N^{\star}(V_s^q) + 3V_v^q ,
\label{efmas}
\end{equation}
where the effective nucleon mass $M_N^{\star}$ depends on {\em only the
scalar potential\/} in the medium.  This important observation was also
obtained by non-local field theory model~\cite{nonloc}, where a non-local
$\sigma$-$\omega$ model containing short distance vertex form factors is
used to simulate an underlying QCD substructure.

Although we have discussed the QMC model using
the specific model, namely the bag model, in our previous
works~\cite{guichon,qmc,qmc2,review1}, 
the qualitative features we have found may be correct in any model in
which the nucleon contains
{\em relativistic} quarks and the (middle- and long-range) attractive
and (short-range) repulsive N-N forces have Lorentz-scalar and
vector characters, respectively.  To confirm this point 
we will study nuclear matter using 
the quark models discussed in the previous section.  

In Eq.(\ref{dirac2}) the scalar and vector potentials are respectively
mediated by the $\sigma$ and $\omega$ mesons.   Their
mean-field values are introduced by $V_s^q = g_{\sigma}^q \sigma$ and
$V_v^q = g_{\omega}^q \omega$, where
$g_{\sigma}^q$ ($g_{\omega}^q$) is the coupling constant of the
quark-$\sigma$ 
($\omega$) meson.  (We here ignore the $\rho$ meson and the Coulomb field,
for simplicity.  For those fields, see Ref.~\cite{qmc}.)
Then, in MFA the effective Lagrangian density in the QMC model would be
given 
\begin{equation}
{\cal L}_{QMC}= \overline{\psi} [i \gamma \cdot \partial
- M_N^{\star}(\sigma)
- g_\omega \gamma_0 \omega ] \psi
- \frac{1}{2}[ (\nabla \sigma)^2 + m_{\sigma}^2 \sigma^2 ]
+ \frac{1}{2}[ (\nabla \omega)^2 + m_{\omega}^2 \omega^2 ],
\label{qmclag}
\end{equation}
where $\psi$ is the nucleon field, and $m_\sigma$ and $m_\omega$
are respectively 
the masses of the $\sigma$ and $\omega$ mesons --- we take $m_\sigma = 550$
MeV and $m_\omega = 783$ MeV.  The $\omega$-N coupling constant $g_\omega$
is related to the corresponding quark-$\omega$ 
coupling constant as $g_\omega = 3 g_\omega^q$~\cite{qmc}.
In an uniform nuclear matter the derivative terms of the meson fields
vanish because the fields do not depend on time and position.

Here we consider the nucleon mass in matter further.  The nucleon mass is a
function of the scalar field.  Because the scalar field is small
at low density the nucleon mass can be expanded in terms of $\sigma$ 
\begin{equation}
M_N^{\star} = M_N + \left( \frac{\partial M_N^{\star}}{\partial \sigma}
\right)_{\sigma=0} \sigma + \frac{1}{2} \left( \frac{\partial^2 M_N^{\star}}
{\partial \sigma^2} \right)_{\sigma=0} \sigma^2 + \cdots .
\label{nuclm}
\end{equation}
Since the interaction between the quark and the $\sigma$ field is described
in terms of the local scalar coupling, the derivative of $M_N^{\star}$ with
respect to $\sigma$ is given by
\begin{equation}
\left( \frac{\partial M_N^{\star}}{\partial \sigma} \right)
= -3g_{\sigma}^q \int_{V_N} d{\vec r} \ \ {\overline \psi}_q \psi_q
\equiv -3g_{\sigma}^q S_N(\sigma) .
\label{deriv}
\end{equation}
Here we have defined the quark-scalar charge in the nucleon $S_N(\sigma)$,
which is itself a function of the scalar field, by Eq.(\ref{deriv}).
Because $S_N$ is positive, the nucleon mass decreases in matter
at low density.  Note that since the nucleon mass is given by
Eq.(\ref{nmass2}) in the HO model the derivative turns to
\begin{equation}
\left( \frac{\partial M_N^{\star}}{\partial \sigma} \right)
= -2g_{\sigma}^q S_N(\sigma) .
\label{hoderiv}
\end{equation}

We furthermore define the scalar-charge ratio, $S_N(\sigma)/S_N(0)$,
to be $C_N(\sigma)$ and the $\sigma$-N coupling constant $g_\sigma$:
\begin{equation}
C_N(\sigma) = S_N(\sigma)/S_N(0) \ \ \mbox{and} \ \
g_{\sigma} = 3g_{\sigma}^q S_N(0) .
\label{cn}
\end{equation}
Using those quantities, we find that the nucleon mass in matter is
\begin{equation}
M_N^{\star} = M_N - g_{\sigma} \sigma - \frac{1}{2} g_{\sigma}
C_N^\prime(0) \sigma^2 + \cdots .
\label{nuclm2}
\end{equation}
(In the HO model, we find $M_N^{\star} = M_N - \frac{2}{3}g_{\sigma}
\sigma - \frac{1}{3} g_{\sigma}C_N^\prime(0) \sigma^2 + \cdots$.)
In general, $C_N$ is a decreasing function because the quark in matter is
more relativistic than in free space (because of the attractive force due to
the $\sigma$).    Thus, $C_N^\prime(0)$ takes a
negative value. If the nucleon were structureless $C_N$ would not depend on
the scalar field, that is, $C_N$ would be constant ($C_N=1$).  Therefore,
only the first two terms in the right hand side of Eq.(\ref{nuclm2}) remain,
which is exactly the same as the equation for the effective nucleon
mass in the original QHD model~\cite{qhd}.

\subsection{Numerical results}
\label{subsec:numerical}

We consider an isosymmetric nuclear matter with Fermi momentum
$k_F$, which is given by $\rho_B = 2 k_F^3 / 3\pi^2$.  From the 
Lagrangian density Eq.(\ref{qmclag}), the total energy per nucleon 
$E_{tot}$ can be written
\begin{equation}
E_{tot} = \frac{4}{\rho_B (2\pi)^3} \int^{k_F}
d\vec{k} \sqrt{M_N^{\star 2} + \vec{k}^2} + \frac{m_{\sigma}^{2}}
{2\rho_B}{\sigma}^2 + \frac{g_{\omega}^2}
{2m_{\omega}^{2}}\rho_B ,
\label{tote}
\end{equation}
where the effective nucleon mass in matter is calculated by the quark
model.  The value of the $\omega$ field is determined by baryon number
conservation as $\omega =g_{\omega}\rho_B / m_{\omega}^{2}$~\cite{qmc}.

The scalar mean-field is given by a self-consistency
condition~\cite{qmc}
\begin{eqnarray}
{\sigma} &=& - \frac{4}{(2\pi)^3 m_{\sigma}^{2}}
\int^{k_{F}} d\vec{k} \frac{M_N^{\star}}
{\sqrt{M_N^{\star 2} + \vec{k}^2}} \left(\frac{\partial
M_N^{\star}}{\partial {\overline \sigma}}\right), \nonumber \\
&=& \frac{4g_\sigma C_N(\sigma)}{(2\pi)^3 m_{\sigma}^{2}}
\int^{k_{F}} d\vec{k} \frac{M_N^{\star}}
{\sqrt{M_N^{\star 2} + \vec{k}^2}} ,
\label{scc}
\end{eqnarray}
where the derivative of the effective nucleon mass is given by the
quark-scalar charge $S_N(\sigma)$.  (See Eqs.(\ref{deriv}) and (\ref{cn}).
For the HO model we should replace $C_N$ by $\frac{2}{3}C_N$.)

In the previous section 
we have prepared the relativistic constituent quark model and the
MIT bag model to describe the nucleon. 
We use $m_q$ = 300 MeV in the constituent quark model, while the quark is 
assumed to be massless in the bag model.  In the SW model
we set the radius of the potential to be $R = 0.8$ fm and detrmine $z$
so as to fit the free nucleon mass $M_N$ (= 939 MeV).
The parameter $\beta$ is chosen to be 0 (the confining potential is
purely Lorentz-scalar), 0.25 and 0.5 to examine the effect of
Lorentz-vector type confining potential.  The quark eigenvalue
$x$ can be found by solving the boundary condition Eq.(\ref{bound}) at the
potential surface.  In the HO model, there are two adjustable parameters,
$c$ and $E_s$.  Those are determined so as to fit the free nucleon mass
(see Eq.(\ref{nmass2})) and the root-mean-square (rms) (charge) radius 
of the free nucleon: 
$r_q (= \langle r^2 \rangle^{1/2}) = 0.6$ fm~\cite{rms} (see Appendix).
The quark energy $E$ is given by the condition Eq.(\ref{hocond}).
We found that $c = 1.591$ fm$^{-3}$, $E_s = 344.7$ MeV and $E = 641.8$ MeV 
for the free nucleon.  In nuclear matter we keep $c$ and $E_s$ 
constant and the quark energy $E$ varies depending on the scalar field. 

As in Ref.\cite{qmc}, in the bag model the bag constant $B$ and the
parameter $z$ are fixed to reproduce the free nucleon mass under the
condition that the nucleon mass be stationary under variation of
the bag radius $R$.  As in the SW model, we choose the bag radius of the
free nucleon to be 0.8 fm.  (Variations of the quark mass and the radius
only lead to numerically small changes in the calculated
results~\cite{qmc}.)  We found $B^{1/4}$ = 170.3 MeV for massless quarks.
In Table~\ref{tab:param} we list up the parameter $z$, quark eigenvalue 
and rms radius of the free nucleon in the SW 
and bag models.  

Now we are in a position to determine the coupling constants: 
$g_{\sigma}^2$ and $g_{\omega}^2$ are fixed to fit the binding energy
($-15.7$ MeV) at the saturation density ($\rho_0 = 0.15$ fm$^{-3}$) for
symmetric nuclear matter.  The coupling constants
and some calculated properties for matter are listed in Table~\ref{tab:cc}.
We also present the ratios of the quark eigenvalue, rms radius, axial vector
coupling constant ($g_A$), magnetic moment of the proton ($\mu_p$)
and quark-scalar charge
at saturation density to those in free space.  (See also Ref.~\cite{st}.
For explicit expressions of those quantities, see Appendix.)

We note that the present quark models (except B(0.25) and B(0.5)) 
provide the similar results in the properties of the in-medium nucleon and 
nuclear matter.  In particular, they give values for 
the nuclear compressibility well within the 
experimental range ($K \sim$ 200 -- 300 MeV), which is 
caused by the non-linear dependence of the $\sigma$ field on the
nucleon mass.  In the cases of B(0.25) and B(0.5)
the reduction of the nucleon mass in matter is relatively small, which
is due to the strong Lorentz-vector type confining
potential.  Furthermore, in the HO model the increase of the rms radius
at saturation density is about 10\%, which is large compared
with the observed value~\cite{sick}.

We briefly comment on the rms radius, axial vector coupling constant
and magnetic moment of the nucleon in nuclear matter.  In our
calculation those quantities are calculated using only the (valence)
quark wave function and other important contributions from the meson
(mainly pion) cloud effect etc are not included.  In the present quark
picture those quantities involve integrals over the lower and upper
components of the quark wave function.  Because of the attractive force due
to the scalar field the quark becomes more relativistic in matter than
in free space, that is, the lower component is more enhanced in matter.
This fact gives the decrease in $g_A$ and the increase in the rms radius and
$\mu_p$ in matter.

\begin{figure}[t]
\begin{center}
\epsfig{file=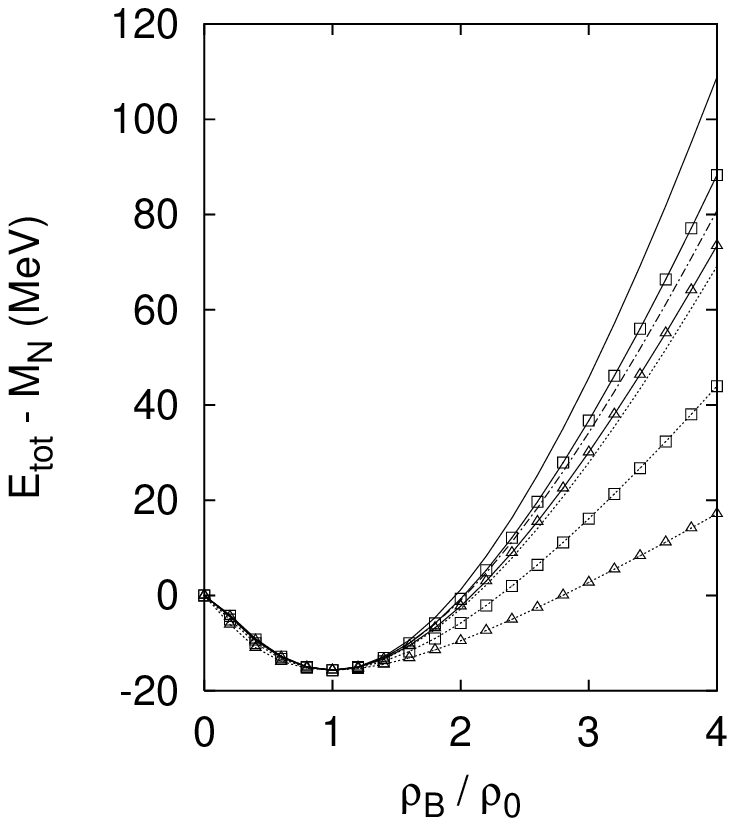,height=10cm}
\caption{Energy per nucleon for symmetric nuclear matter.
The saturation curves for 
the SW model and the bag model are respectively shown by solid and
dotted curves.  The dot-dashed curve presents the result of the HO model.
The curves with squares (triangles) are for the results with $\beta = 0.25
(0.5)$, while the curves without any marks are for $\beta = 0$.
}
\label{fig:etot}
\end{center}
\end{figure}

\begin{figure}[t]
\begin{center}
\epsfig{file=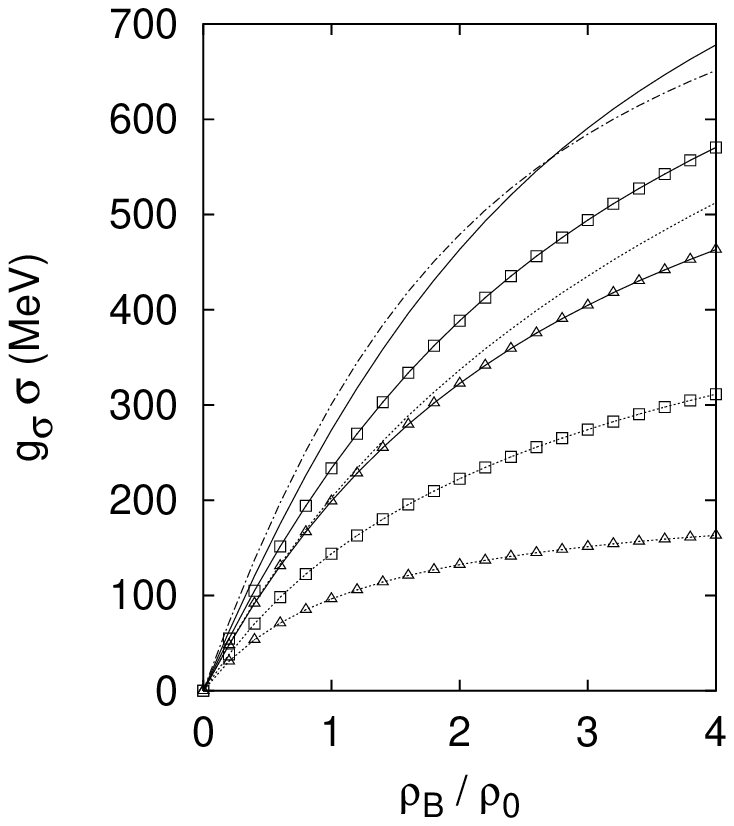,height=10cm}
\caption{Scalar mean-field value.  The curves are labeled as in
Fig.~\protect\ref{fig:etot}. 
}
\label{fig:gsig}
\end{center}
\end{figure}

\begin{figure}[t]
\begin{center}
\epsfig{file=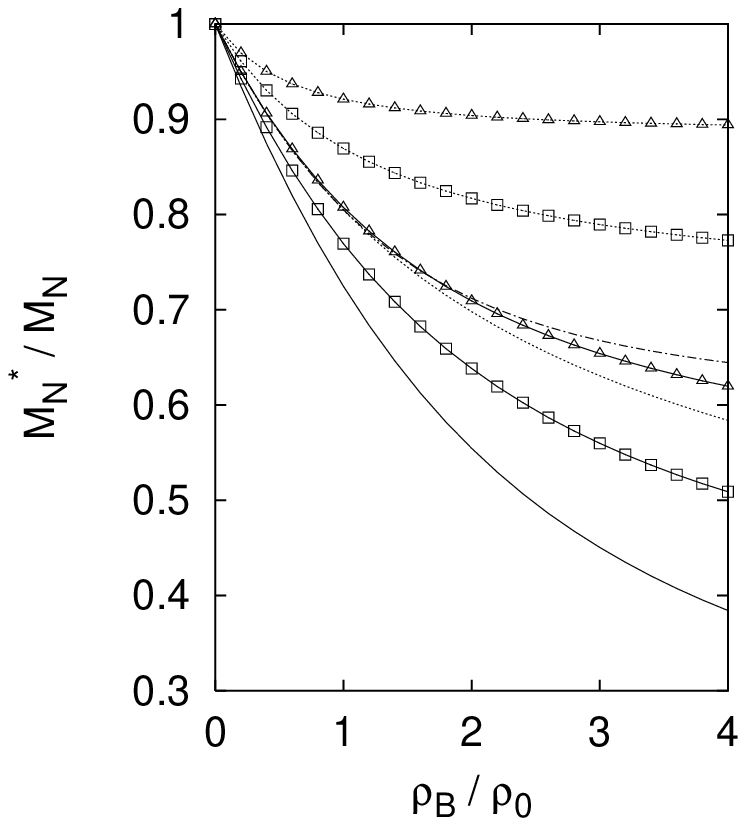,height=10cm}
\caption{Effective nucleon mass in symmetric nuclear matter. The curves
are labeled as in Fig.~\protect\ref{fig:etot}.
}
\label{fig:nmass}
\end{center}
\end{figure}

In Fig.~\ref{fig:etot} we present the total energy per nucleon 
as a function of nuclear density.  Around normal nuclear matter density 
the quark models provide 
the similar saturation curves, while for high density the curves obtained
from the models of B(0.25) and B(0.5) are relatively lower than those of
the others.  This corresponds to the fact that a quark model
with a larger value of $\beta$ gives a lower nuclear compressibility
(see Table~\ref{tab:cc}).  We show the scalar mean-field value and the
effective nucleon mass for symmetric nuclear matter in
Figs.~\ref{fig:gsig} and \ref{fig:nmass}, respectively.  From those figures 
we again see that in the cases of B(0.25) and B(0.5) the scalar field 
is much weaker 
than those given by the other quark models and the decrease of the
nucleon mass is fairly small even at high density, which is due to
the small scalar field. 

We note that in the SW and bag models with
massless quarks the quark-scalar charge vanishes in the limit
$\beta \to 1$, which means that the $\sigma$ meson does not couple to
the nucleon. (For details, see Appendix.)  This fact implies that as
$\beta$ is larger the $\sigma$-N coupling is weaker in matter.
Thus, we may conclude that qualitatively a large
mixture of the Lorentz-vector type confining potential leads to a weak 
scalar mean-field and a large effective nucleon mass in nuclear
matter.  It is well known that within MFA
a smaller effective nucleon mass (and hence a stronger scalar field) in 
matter is favorable to fit various properties of finite nuclei~\cite{qhd}. 
Therefore, the confining potential including a strong 
Lorentz-vector type one may not be suitable for describing a nuclear system.  

\section{Transformation from QMC to QHD}
\label{sec:qhdtype}

The main difference between the QMC and QHD models lies in the
dependence of the scalar field on the nucleon mass in matter.
By performing a re-definition of the scalar field, the QMC Lagrangian
density at hadronic level can be 
cast into a form silimar to a QHD-type mean-field model, in which the
nucleon mass depends on the scalar field linearly, with
self-interactions of the scalar field~\cite{newqhd,muller}.
Here we will briefly review the
field re-definition and discuss, as an example, the derivative (scalar)
coupling model of QHD (i.e., the Zimanyi-Moszkowski (ZM) model)~\cite{zm}.
Then, we will calculate
non-linear scalar potentials generated from the QMC model and
compare with the potentials phenomenologically determined by fitting
the properties of finite nuclei and nuclear matter.

\subsection{Field re-definition}
\label{subsec:redef}

In the QMC model the nucleon mass in matter is given by a function
of the scalar field $\sigma$ 
\begin{equation}
M_{N,QMC}^{\star} = M_{N,QMC}^{\star}(\sigma) ,
\label{effnmas}
\end{equation}
while in QHD the mass depends on a scalar field $\phi$ linearly, 
\begin{equation}
M_{N,QHD}^{\star} = M_{N} - g_0 \phi .
\label{effnmas2}
\end{equation}
This fact suggests a re-definition of the scalar field in QMC:
\begin{equation}
g_0 \phi(\sigma) = M_N - M_{N,QMC}^{\star}(\sigma) ,
\label{redef}
\end{equation}
where $g_0$ is a constant chosen so as to normalize the new scalar
field $\phi$ in the limit $\phi \to 0$:
\begin{equation}
\phi(\sigma) = \sigma + {\cal O}(\sigma^2) .
\label{newnorm}
\end{equation}
Thus, $g_0$ is given by
\begin{equation}
g_0 = - \left( \frac{\partial M_{N,QMC}^{\star}(\sigma)}{\partial \sigma}
\right)_{\sigma = 0}.
\label{g0}
\end{equation}
Comparing with Eqs.(\ref{deriv}) and (\ref{cn}), we find $g_0 = g_\sigma$
for the SW and bag models and $g_0 = \frac{2}{3}g_\sigma$ for the HO model.

The contribution of the scalar field to the total energy, $E_{scl}$, is now 
rewritten in terms of the new field $\phi$.  From Eq.(\ref{qmclag}) we find
\begin{equation}
E_{scl} = \frac{1}{2} \int d\vec{r} \ [(\nabla \sigma)^2 + m_{\sigma}^2
\sigma^2 ]
= \int d\vec{r} \ [\frac{1}{2}h(\phi)^2(\nabla \phi)^2 + U_s(\phi) ] ,
\label{es}
\end{equation}
where $U_s$ describes the self-interactions of the scalar field
\begin{equation}
U_s(\phi) = \frac{1}{2} m_{\sigma}^2 \sigma(\phi)^2 ,
\label{Us}
\end{equation}
and
\begin{equation}
h(\phi) = \left( \frac{\partial \sigma}{\partial \phi} \right)
= \frac{1}{m_\sigma \sqrt{2U_s(\phi)}}
\left( \frac{\partial U_s(\phi)}{\partial \phi} \right) .
\label{h}
\end{equation}
Note that in uniformly distributed nuclear matter the first derivative
term in $E_{scl}$ does not contribute to the total energy.
(The effect of this term on the properties of finite nuclei was
studied in Ref.~\cite{muller}.)

The QMC model is now re-formulated in terms of the new scalr field $\phi$,
and it is of the same form as a QHD model with the non-linear scalar
potential $U_s(\phi)$, which is generated from the effect of the internal
structure of 
the nucleon, and with the coupling $h(\phi)$ to the gradient of the scalar 
field.  However, we should keep in mind that this re-formulation is
performed only at hadronic level and that the non-linear potential $U_s$
is simply determined so as to simulate the effective nucleon mass given by
the quark model in matter. 
Thus, {\em only on the energy of nuclear matter}
the QHD-type mean-field model derived here is identical to the QMC model.
We will again discuss this point in the last section.

The standard form of the non-linear scalar potential
is~\cite{boguta,review2}
\begin{equation}
U_s(\phi) = \frac{1}{2} m_{\sigma}^2 \phi^2 + \frac{\kappa}{6} \phi^3
+ \frac{\lambda}{24} \phi^4 .
\label{stUs}
\end{equation}
In more sophisticated version of QHD~\cite{newqhd}, inspired by modern
methods of effective field theory, many other terms of meson-meson and
meson-nucleon couplings are considered.  However, rather than 
attempting to compete with those models, in this paper we concentrate only on 
the parameters $\kappa$ and $\lambda$ to make our discussion simple. 
It is well known that the non-linear scalar potential Eq.(\ref{stUs}) 
is practically indispensable to reproduce the bulk properties of finite 
nuclei and nuclear matter~\cite{boguta,review2,newqhd,zm,nl1,tm1}.

If the effective nucleon mass could be determined experimentally 
as a function of the scalar 
field, this procedure leading to Eq.(\ref{Us}) would
allow to predict a part of the non-linear potential, which is
originated by the quark substructure of the nucleon in nuclear matter.
Inversely, if the non-linear potential obtained by fitting various
experimental data could be separated into a portion due to 
the effect of the internal structure of the nucleon and the inherent
self-interaction part of the scalar field, we could get significant
information on the quark structure of the nucleon in matter by using
the inverse field transformation,
\begin{equation}
\sigma(\phi) = \frac{\sqrt{2U_s(\phi)}}{m_\sigma}.
\label{inverse}
\end{equation}
It is potentially of very interesting.  However, it may be quite difficult.

As an example, we here consider the derivative coupling model
of QHD, namely the Zimanyi-Moszkowski (ZM) model~\cite{zm}.  By
re-definig the scalar field the ZM model can be exactly transformed to
a QHD-type model with a non-linear potential.

In the ZM model the interaction term between the $\sigma$ and the nucleon
is given by~\cite{zm} 
\begin{equation}
{\cal L}_{int}^{ZM}= (g_\sigma \sigma /M_N)
\overline{\psi} i \gamma \cdot \partial \psi ,
\label{zmint}
\end{equation}
in addition to a similar coupling to the vector field.  Then, after
rescaling the nucleon field, the Lagrangian density turns to be the
same form as that of the original QHD, with the effective nucleon mass
\begin{equation}
M_N^\star = \frac{M_N}{1+(g_\sigma \sigma /M_N)},
\label{zmmass}
\end{equation}
instead of the mass depending on the scalar field linearly.  
Thus, the model involves higher order 
couplings between the $\sigma$ and the nucleon, which provides
a good nuclear compressibility around 230 MeV and $M_N^\star /M_N
\simeq 0.85$~\cite{zm}.

Let us define a new scalar field $\phi$ by 
\begin{equation}
g_0 \phi(\sigma) = M_N - M_{N,ZM}^{\star}(\sigma) ,
\label{zmredef}
\end{equation}
where the mass $M_{N,ZM}^{\star}(\sigma)$ is given by Eq.(\ref{zmmass}).
Then, we easily find
\begin{equation}
\phi(\sigma) = \frac{\sigma}{1 + g^\prime \sigma} ,
\label{zmphi}
\end{equation}
with $g^\prime = g_\sigma /M_N$.  Inversely, we also find
\begin{equation}
\sigma(\phi) = \frac{\phi}{1 - g^\prime \phi} .
\label{zmsigma}
\end{equation}
The non-linear potential is thus given by Eq.(\ref{Us})
\begin{equation}
U_s(\phi) = \frac{1}{2} m_{\sigma}^2 \sigma(\phi)^2
= \frac{1}{2} m_{\sigma}^2 \left( \frac{\phi}{1 - g^\prime \phi} \right)^2 ,
\label{zmUs}
\end{equation}
which can be expanded up to order ${\cal O}(\phi^4)$
\begin{equation}
U_s(\phi) \simeq \frac{1}{2} m_{\sigma}^2 \phi^2
+ g_\sigma \left( \frac{m_{\sigma}^2}{M_N} \right) \phi^3
+ \frac{3}{2} g_\sigma^2 \left( \frac{m_{\sigma}}{M_N} \right)^2 \phi^4 .
\label{zmstUs}
\end{equation}
(The shape of the potential Eq.(\ref{zmUs}) is shown in
Fig.\ref{fig:potrmf}.)
Hence, the ZM model is identical to the QHD-type model with the
non-linear scalar potential expressed by Eq.(\ref{zmUs}).

\subsection{Non-linear scalar potentials generated from QMC}
\label{subsec:nonpot}

In general the nucleon mass in matter may be given by a complicated
function of the scalar field.  However, in QMC the mass can be
parametrized by a simple expression up to
${\cal O}(g_\sigma^2)$~\cite{qmc,qmc2}  (see Fig.~\ref{fig:nmass})
\begin{equation}
\frac{M_N^\star}{M_N} \simeq 1 - a y + b y^2 ,
\label{mparam}
\end{equation}
with a dimensionless scale $y = g_\sigma \sigma / M_N$ and
two (dimensionless) parameters $a$ and $b$. From the discussion
around Eq.(\ref{nuclm2}) we can expect that $a \sim 1$ for the SW and
bag models while $a \sim 2/3$ for the HO model.  This parametrization
is accurate up to $\sim 4\rho_0$.
Using Eq.(\ref{mparam}) the calculated nucleon mass in
nuclear matter can be fitted with the method of least squares.  We show
the parameters $a$ and $b$ for the various quark models
in Table~\ref{tab:ab}.  We find that the present quark models (except 
B(0.25) and B(0.5)) give $b \sim 0.2 - 0.5$.  

Now we shall re-define the scalar field using Eq.(\ref{redef}).  
We find $g_0 = a g_\sigma$ and
\begin{equation}
\phi(\sigma) = \sigma - d \sigma^2 ,
\label{qmcphi}
\end{equation}
with $d = bg_\sigma /aM_N$.
Eq.(\ref{qmcphi}) can be solved exactly, and we obtain
\begin{equation}
\sigma(\phi) = \frac{1-\sqrt{1-4d\phi}}{2d} ,
\label{qmcsig}
\end{equation}
which satisfies the condition $\sigma \to 0$ in the limit
$\phi \to 0$.  The non-linear potential is then calculated
\begin{equation}
U_s(\phi) = \frac{m_\sigma^2}{2} \left( \frac{\sigma(\phi) - \phi}{d}
\right) , 
\label{qmcnonl}
\end{equation}
and we can expand $U_s$ as 
\begin{equation}
U_s(\phi) = \frac{m_\sigma^2}{2} \phi^2
+ g_\sigma r \left( \frac{m_\sigma^2}{M_N} \right) \phi^3
+ \frac{5}{2} g_\sigma^2 r^2 \left( \frac{m_\sigma}{M_N} \right)^2 \phi^4
+ {\cal O}(g_\sigma^3) ,
\label{qmcnonlex}
\end{equation}
where $r = b/a$.

Comparing with Eq.(\ref{stUs}) we can find the
parameters $\kappa$ and $\lambda$ in the non-linear scalar potential
generated by QMC.  They are listed in Table~\ref{tab:kl}.
The quark models we have adopted here lead to the non-linear potential with
$\kappa \sim 20 - 40$ (fm$^{-1}$) and $\lambda \sim 80 - 400$, except for
the models of B(0.25) and B(0.5).  (The models B(0.25) 
and B(0.5) give quite large values of $\kappa$ and $\lambda$, in
addition to the large effective nucleon mass and the small scalar field
in matter, which seems impractical.)  From the discussion in the 
subsection~\ref{subsec:nucleonstr} we see that both $a$ and $b$ are 
positive in QMC because $a$ and $b$ are respectively proportional to
$S_N(0)$ and $-C_N^\prime(0)$.  Hence, from Eq.(\ref{qmcnonlex}) we can
expect that the effect of the quark substructure of the nucleon in matter is
cast into a non-linear potential with {\em positive} $\kappa$ and
{\em positive} $\lambda$ in the QHD-type mean-field model.

\begin{figure}[t]
\begin{center}
\epsfig{file=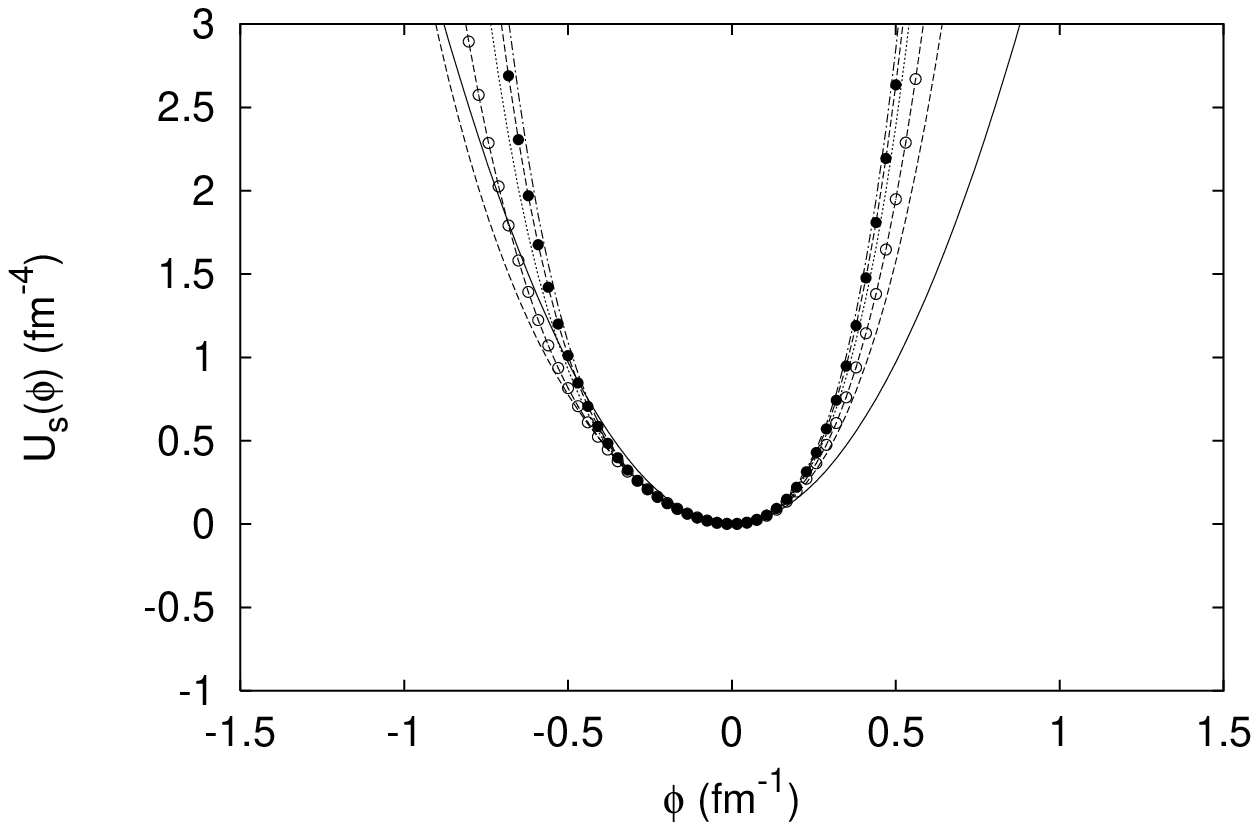,height=10cm}
\caption{Non-linear scalar potentials generated from the QMC model.
The solid curve shows $U_s = \frac{m_\sigma^2}{2} \phi^2$.
The dashed curve with open (solid) circles is for the SW 
model with $\beta = 0.25 (0.5)$, while the dashed one without any marks
is for the SW model with $\beta = 0$.  The result of the B(0) model is
shown by the dotted curve.  The dot-dashed curve is for the HO model.
}
\label{fig:potqmc}
\end{center}
\end{figure}

\begin{figure}[t]
\begin{center}
\epsfig{file=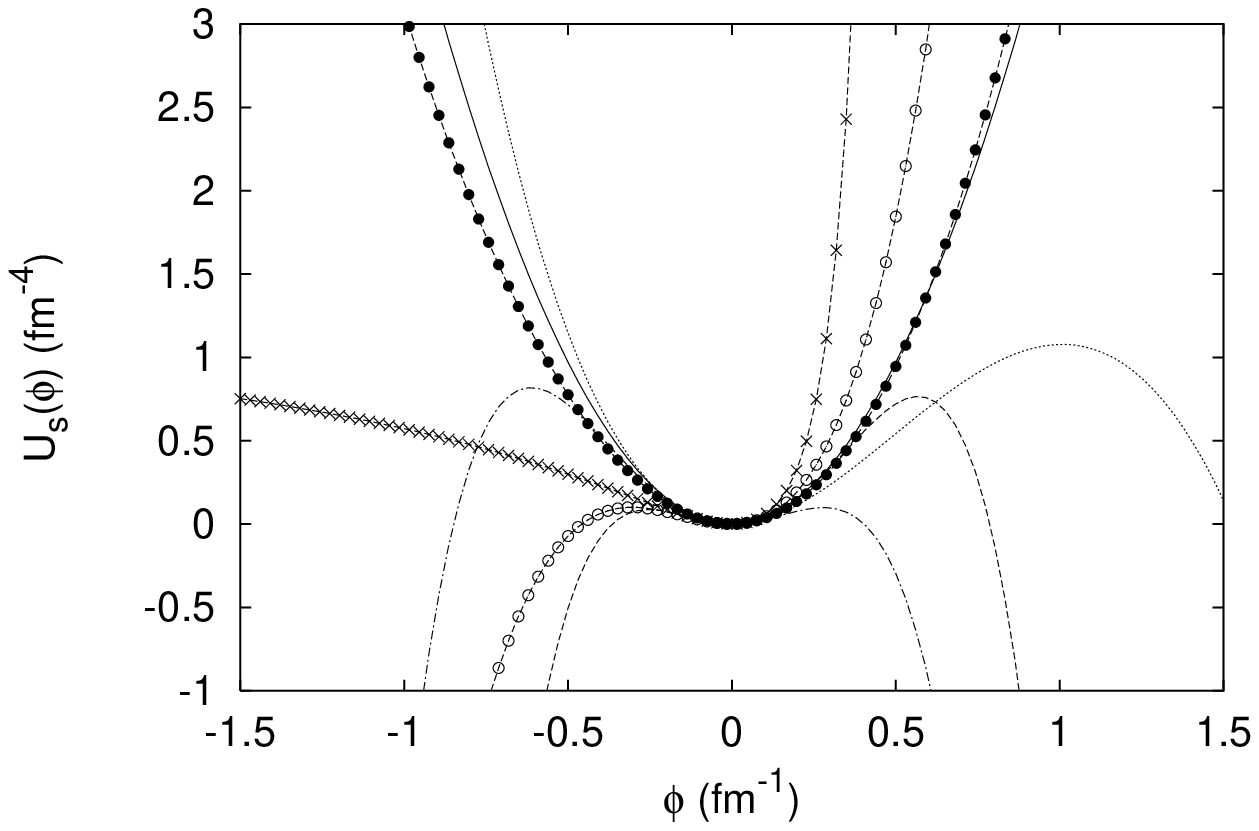,height=10cm}
\caption{Non-linear scalar potentials in the RMF models.
The solid curve shows $U_s = \frac{m_\sigma^2}{2} \phi^2$.
The dashed curve with open (solid) circles is for the G2
(NLB) model~\protect\cite{newqhd}, while the dashed one without any marks
is for the G1 model~\protect\cite{newqhd}.  The dashed curve with crosses
is for the ZM model~\protect\cite{zm}.  The potentials in
the TM1~\protect\cite{tm1} and NL1~\protect\cite{nl1} models are
respectively shown by the dotted and dot-dashed curves.
}
\label{fig:potrmf}
\end{center}
\end{figure}

In Table~\ref{tab:klrmf}, for comparison we show the parameters
$\kappa$ and $\lambda$, which are phenomenologically determined 
in various relativistic mean-field (RMF) models.  In such 
models $\kappa$ and $\lambda$ are sometimes taken to be
negative.  Furthermore, in Figs.~\ref{fig:potqmc} and \ref{fig:potrmf}
we respectively illustrate the shapes of the non-linear scalar potentials
calculated from the QMC model and those in the RMF 
models.  (In Fig.~\ref{fig:potqmc} we did not show the results of B(0.25) and 
B(0.5).)  From Fig.~\ref{fig:potqmc} 
we can see that the various quark models lead to the similar 
non-linear potentials, 
despite of the big difference in the confinement mechanism.  
It may imply that 
we cannot discriminate the quark models for the nucleon only from the
point of view of the energy of nuclear matter.  On the other hand,
the non-linear potentials in the RMF models show quite different behaviors
in Fig.~\ref{fig:potrmf}.  In the region of $\phi>0$ the potentials
in the G2, NLB and ZM models are close to those produced by QMC.

We here comment on the coupling $h(\phi)$ to the gradient of the scalar
field.  The QMC model gives
\begin{equation}
h(\phi) = 1 + g_{s1} \left( \frac{\phi}{M_N} \right)
+ g_{s2} \left( \frac{\phi}{M_N} \right)^2
+ {\cal O}(g_\sigma^3) ,
\label{qmch}
\end{equation}
where $g_{s1} = 2 g_\sigma r$ and $g_{s2} = 6 g_\sigma^2 r^2$.  We find
that the present model suggests $g_{s1} \sim 4 - 8$ and 
$g_{s2} \sim 20 - 100$, while the ZM model gives $g_{s1} \sim 30$ and 
$g_{s2} \sim 580$.  In the RMF model, for example, the G1 and G2
models lead to $g_{s1} \sim 9$, which is consistent with the values
obtained in QMC.  This coupling $h(\phi)$ affects the surface properties
of finite nuclei~\cite{muller}.

\section{Discussion and conclusion}
\label{sec:concl}

We have first studied the MIT bag model and the relativistic constituent
quark model for the nucleon, where a square well and harmonic oscillator
potentials are used to confine the quarks.  We have
considered not only the Lorentz-scalar type confining potential but also
the Lorentz-vector type one in order to study the role of the vector 
potential in nuclear matter.  Then, we have calculated the properties of  
the nucleon and nuclear matter using QMC with those quark models.  
As a mixture of the Lorentz-vector 
type confining potential is larger the effective nucleon mass in matter
is larger and the strength of the scalar field is weaker.  In particular, 
it is preferable in the bag model that the Lorentz structure of the
confinement is purely scalar (or including a very weak Lorentz-vector type
potential). 

Next, we have performed a re-definition of the scalar field in matter and 
transformed the QMC model to a QHD-type model with a non-linear scalar
potential.  We then compared our results with the potentials which are
determined so as 
to fit the properties of finite nuclei and nuclear matter in the
RMF models.  The QMC model provides the parameters $\kappa \sim 20 - 40$
(fm$^{-1}$) and $\lambda \sim 80 - 400$ for the non-linear scalar potential.
The shapes of the potentials generated from QMC are thus very
close to one another, despite of the different
confinement mechanism in the quark models.  On the other hand,
the parameters phenomenologically determined 
in the RMF models take various values and the shapes of the potentials are
quite different.  In general, the phenomenological potential may consist of
a part, which is caused by the quark substructure of the nucleon, and
the inherent self-couplings of the scalar field.  Therefore, if 
the part due to the internal structure of the nucleon could be known by 
nuclear experiments, we could get significant 
information on the quark structure of the in-medium nucleon. 

We should emphasize that the field re-definition performed here enables
us to make a QHD-type effective Lagrangian from the QMC model.  However,
the created Lagrangian is identical to the QMC model {\em only on
the energy of a nuclear system}.  In QMC the quark wave function in the
nucleon is modified by the nuclear environment self-consistently at each
nuclear density.  We can know it explicitly.  It is the big difference
between the QHD-type and QMC models, and is of great advantage to
the QMC model.  Using the change of the quark wave function in the medium 
provided 
by QMC we can explain many intriguing nuclear phenomena~\cite{review1}, for
example, the nuclear EMC effect~\cite{emc2}, charge symmetry breaking in
nuclei (the Okamoto-Nolen-Schiffer anomaly)~\cite{ons}, change of
electromagnetic form factors of the bound nucleon~\cite{form}, longitudinal
response functions in electron scattering~\cite{resp} etc.

We have one comment on the modified quark-meson coupling (MQMC) model 
proposed by Jin and Jennings~\cite{jin}.  
The original QMC model~\cite{guichon,qmc,qmc2} generates 
a (relatively) small scalar field in nuclear matter.
On the other hand, it is well
known that the MQMC model can generate the scalar field, which is close to
that in QHD (see also Ref.~\cite{muller}).  However, unfortunately 
the increase of the nucleon size at $\rho_0$ is about 40\%
of the free-space value, which is apparently too large~\cite{sick,form}. 
Although it must be true that the bag constant decreases
as the density grows up, the reduction of the bag constant around
normal nuclear matter density may not be large.  For example,
the Dyson-Schwinger approach at finite chemical
potential~\cite{bender} suggested that the bag constant is reduced very
little at $\rho_0$, while it decreases very rapidly near the phase
transition point.  

It is possible to extend the present QMC model to
a model which includes quark degrees of freedom in the mesons, as well as
in the nucleons --- we call it QMC-II~\cite{qmc2,natural}. 
The QMC-II model can provide a lot of effective coupling terms among
the meson fields because the mesons have structure.
In particular, the QMC-II Lagrangian automatically offers non-linear
terms with respect to the meson fields.  Those interaction terms may
correspond to the terms of meson-meson and meson-nucleon couplings
considered in the new version of QHD~\cite{newqhd} or higher order
terms appearing in the chiral effective lagrangian for nuclear
matter~\cite{gel}.  
It will be very interesting to construct a QMC model, in which both the
quark substructure of hadrons and non-linear
potentials due to inherent self-couplings of the meson fields
are involved (see, for example, Ref.~\cite{toki}), and 
explore the connection between various coupling strengths found 
empirically in the phenomenological models and those predicted in 
the QMC model. 

\newpage
\begin{center}
{\bf APPENDIX}
\end{center}

In the relativistic quark models we adopt here 
the lowest energy state for the confined quark is given by Eq.(\ref{qsol}).
In the SW model the upper and lower components of the quark wave function 
inside the well are respectively given by
the spherical bessel functions: $f(r) = j_0(kr)$ and
$g(r) = \sqrt{(\alpha-\lambda)/(\alpha+\lambda)} j_1(kr)$, where
$\alpha = RE = \sqrt{x^2+\lambda^2}$, $\lambda = m_qR$ and $k = x/R$.
The normalization constant is then given by
\begin{equation}
N^2 = \frac{x^2(1-\beta)}{R^3j_0^2(x)D_{SW}} ,
\label{norm1}
\end{equation}
where $D_{SW} = 2\alpha(\alpha-\sqrt{1-\beta^2}+\beta\lambda) +
\lambda\sqrt{1-\beta^2}$.

Using those quantities the quark-scalar charge $S_N$ is calculated
\begin{eqnarray}
S_N(m_q) &=& \int_{V_N} d{\vec r} \ \ {\overline \psi}_q \psi_q,  \nonumber
\\
&=& \frac{2\lambda(\alpha-\sqrt{1-\beta^2}+\beta\lambda)
+ \alpha\sqrt{1-\beta^2}}
{D_{SW}} .
\label{scharge1}
\end{eqnarray}
We note that $S_N$ vanishes if $m_q = 0$ MeV and $\beta \to 1$, which may be
accidental in the SW and bag models.  In this case there is no coupling
between the $\sigma$ meson and the nucleon, which implies that the nuclear
matter cannot be described by this model.

The root-mean-square (rms) (charge) radius of the nucleon is given by
\begin{equation}
r_q^2 \equiv \langle r^2 \rangle =
\int_{V_N} d{\vec r} \ \ r^2 \psi_q^\dagger  \psi_q
= R^2 \frac{N_{rms}}{3x^2D_{SW}},
\label{rms1}
\end{equation}
where
\begin{eqnarray}
N_{rms} &=& \alpha [2x^2(\alpha-\sqrt{1-\beta^2})
+ 2\alpha(2+\beta)-3] - 3\lambda \left[\lambda(1+2\beta)-x^2\sqrt{1-\beta^2}
-\frac{3}{2} \right] \nonumber \\
&+& 2\alpha \lambda \left[\left(x^2 - \frac{3}{4}\right) \beta -2\right] .
\label{numrms}
\end{eqnarray}
The axial vector coupling constant ($g_A$) and the magnetic moment
($\mu$) of the nucleon are also calculated analytically.  In the
present quark model $g_A$ is given by $g_A = (5/3) \times
g_A^0$, where the factor $(5/3)$ is the expectation value of
the spin-isospin operator with respect to the nucleon state.  Then,
$g_A^0$ is
\begin{equation}
g_A^0 = \frac{2\alpha^2+4\alpha\lambda-3\lambda\sqrt{1-\beta^2}
+2\beta\lambda(\alpha+2\lambda)}{3D_{SW}}.
\label{ga1}
\end{equation}
The magnetic moment is given by ${\vec \mu} = \sum_i
\mu_i {\vec \sigma}_i Q_i$, where $Q_i$ is the quark charge operator and
the sum $i$ runs over all quarks in the nucleon.  Then, $\mu_i$ is
\begin{equation}
\mu_i = \frac{R}{6}\cdot\frac{4\alpha+2\lambda-3\sqrt{1-\beta^2}
+2\beta(\alpha+2\lambda)}{D_{SW}}.
\label{mu1}
\end{equation}
In the MIT bag model, those quantities are given by the same expressions as
in the SW model.

In the HO model, the quark wave function is again given by Eq.(\ref{qsol}).
The upper and lower components are respectively given by
gaussian functions: $f(r) = e^{-r^2/2r_0^2}$ and
$g(r) = (r/\xi r_0^2) e^{-r^2/2r_0^2}$, where
$r_0 = (c\xi)^{-1/4}$ with $\xi = E + m_q$.  Then, the normalization 
constant is
\begin{equation}
N^2 = \frac{8}{\sqrt{\pi}}\cdot\frac{\xi^2}{D_{HO}} ,
\label{norm2}
\end{equation}
where $D_{HO} = 2(\xi r_0)^2 +3$.

Using the quark wave function, the quark-scalar charge is calculated
\begin{equation}
S_N(m_q) = \frac{2(\xi r_0)^2 - 3}{D_{HO}} .
\label{scharge2}
\end{equation}
We note that $S_N$ approaches $1/3$ when $m_q = 0$ MeV
(and $\beta = 1$).  Similarly,
$g_A^0$ and $\mu_i$ are respectively given by 
\begin{eqnarray}
g_A^0 &=& \frac{2(\xi r_0)^2 - 1}{D_{HO}},
\label{ga2}  \\
\mu_i &=& \frac{\xi r_0^2}{D_{HO}}.
\label{mu2}
\end{eqnarray}

When the nucleon is embedded in a nuclear medium, those quantities can be
expressed by the same forms with the effective quark mass $m_q^{\star}$
(see below Eq.(\ref{dirac2})), instead of the free quark mass $m_q$.


\newpage
\begin{table}
\begin{center}
\caption{Parameter $z$, quark eigenvalue $x$ and rms radius $(r_q)$. 
The bag model is denoted by $B$.  The value of $\beta$ is 
shown in the parentheses.  
}
\label{tab:param}
\begin{tabular}[t]{ccccccc}
\hline
 & SW(0) & SW(0.25) & SW(0.5) & B(0) & B(0.25) & B(0.5) \\
\hline
$z$ & 4.396 & 4.792 & 5.164 & 3.273 & 3.974 & 4.637 \\
$x$ & 2.449 & 2.596 & 2.732 & 2.043 & 2.276 & 2.498 \\
$r_q$ (fm) &
  0.5415 & 0.5264 & 0.5137 & 0.5832 & 0.5721 & 0.5593 \\
\hline
\end{tabular}
\end{center}
\end{table}
%

%
\begin{table}
\begin{center}
\caption{Coupling constants and calculated properties of the nucleon and 
symmetric nuclear matter at normal nuclear density. 
(The quantities with $\star$ are those quantities calculated at $\rho_0$.)
The nuclear compressibility $K$ is quoted in MeV.
The last row is for QHD~\protect\cite{qhd}.
}
\label{tab:cc}
\begin{tabular}[t]{cccccccccc}
\hline
 & $g_{\sigma}^2$ & $g_{\omega}^2$ & $M_N^{\star}/M_N$ & $K$ &
$x^{\star}/x$ & $ r_q^{\star}/r_q$ & $g_A^{\star}/g_A$ &
$\mu_p^{\star}/\mu_p$ & $S_N^{\star}/S_N$ \\
\hline
SW(0) & 84.41 & 104.1 & 0.725 & 329 & 0.937 & 1.03 & 0.936 & 1.11 & 0.886 \\
SW(0.25) & 74.75 & 83.16 & 0.769 & 307 & 0.956 & 1.03 & 0.919 & 1.11 & 0.850
\\
SW(0.5) & 66.61 & 65.21 & 0.807 & 287 & 0.971 & 1.03 & 0.903 & 1.10 & 0.811
\\
HO(1.0) & 146.5 & 64.51 & 0.805 & 309 & --- & 1.10 & 0.891 & 1.23 & 0.812 \\
B(0) & 67.55 & 66.10 & 0.805 & 278 & 0.837 & 1.02 & 0.909 & 1.08 & 0.814 \\
B(0.25) & 54.65 & 35.49 & 0.869 & 235 & 0.900 & 1.03 & 0.877 & 1.06 & 0.711
\\
B(0.5) & 47.99 & 10.92 & 0.921 & 169 & 0.941 & 1.02 & 0.841 & 1.05 & 0.541
\\
\hline
QHD & 91.64 & 136.2 & 0.556 & 540 & --- & --- & --- & --- & --- \\
\hline
\end{tabular}
\end{center}
\end{table}

\newpage
\begin{table}
\begin{center}
\caption{Parameters $a$ and $b$.  Note that $a \sim 1$ for the SW and bag
models, while $a \sim 2/3$ for the HO model.
}
\label{tab:ab}
\begin{tabular}[t]{cccccccc}
\hline
 & SW(0) & SW(0.25) & SW(0.5) & HO(1.0) & B(0) & B(0.25) & B(0.5) \\
\hline
$a$ & 1.008 & 1.006 & 1.017 & 0.6868 & 0.9982 & 1.005 & 0.9933 \\
$b$ & 0.2146 & 0.3229 & 0.4965 & 0.2452 & 0.4346 & 0.9666 & 2.206 \\
\hline
\end{tabular}
\end{center}
\end{table}

\begin{table}
\begin{center}
\caption{Parameters $\kappa$ and $\lambda$ generated by the QMC model.
}
\label{tab:kl}
\begin{tabular}[t]{cccccccc}
\hline
 & SW(0) & SW(0.25) & SW(0.5) & HO(1.0) & B(0) & B(0.25) & B(0.5) \\
\hline
$\kappa$ (fm$^{-1}$)& 19.16& 27.20& 39.03& 42.31& 35.05& 69.65& 150.7\\
$\lambda$ & 78.78 & 158.7 & 326.8 & 384.1 & 263.5 & 1041 & 4872 \\
\hline
\end{tabular}
\end{center}
\end{table}

\begin{table}
\begin{center}
\caption{Parameters $\kappa$ and $\lambda$ for the RMF models.
}
\label{tab:klrmf}
\begin{tabular}[t]{ccccccc}
\hline
 & ZM~\protect\cite{zm} & NL1~\protect\cite{nl1} & TM1~\protect\cite{tm1}
& NLB~\protect\cite{newqhd} & G1~\protect\cite{newqhd}
& G2~\protect\cite{newqhd} \\
\hline
$\kappa$ (fm$^{-1}$) & 74.8& $-24.3$& $-14.5$& 4.06& 29.4& 46.0 \\
$\lambda$ & 720 & $-218$ & 3.71 & 10.0 & $-279$ & 19.7 \\
\hline
\end{tabular}
\end{center}
\end{table}
%


\newpage

\newpage

\newpage

\newpage

\newpage

\end{document}